\documentclass{moriond}

\usepackage{amsmath}
\usepackage{subcaption}

\def\Journal#1#2#3#4{{#1} {\bf #2}, #3 (#4)}

\def\NIM{\em Nucl. Instrum. Methods}
\def\NIMA{{\em Nucl. Instrum. Methods} A}

\def\PLA{{\em Phys. Lett.}  A}

\def\PRL{\em Phys. Rev. Lett.}
\def\PRA{{\em Phys. Rev.} A}
\def\PRD{{\em Phys. Rev.} D}

\def\be{\begin{equation}}
\def\ee{\end{equation}}
\def\bea{\begin{eqnarray}}
\def\eea{\end{eqnarray}}

\newcommand{\Hbar}{\mathrm{\overline{H}}}
\newcommand{\Hbarplus}{\mathrm{\overline{H}^+}}
\newcommand{\pbar}{\mathrm{\overline{p}}}
\newcommand{\positron}{\mathrm{e^+}}
\newcommand{\electron}{\mathrm{e^-}}
\newcommand{\Ps}{\mathrm{Ps}}
\newcommand{\gbar}{\overline{g}}
\newcommand{\BeP}{\mathrm{Be^+}}
\newcommand{\MeV}{\mathrm{MeV}}
\newcommand{\keV}{\mathrm{keV}}

\begin{document}
\vspace*{4cm}
\title{Status and plans of the Gravitational Behaviour of Antihydrogen at Rest experiment}

\author{PHILIPP BLUMER on behalf of the GBAR collaboration}

\address{ETH Zurich, Institute for Particle Physics and Astrophysics \\ Otto-Stern-Weg 5, 8093 Zurich, Switzerland}

\maketitle
\abstracts{The GBAR experiment aims to directly test the Weak Equivalence Principle of ultracold antihydrogen in Earth's gravitational field. The gravitational acceleration $\gbar$ will be measured to a precision of $1\,\%$ using a classical free fall of the anti-atoms from a fixed height. Reaching a precision of $10^{-6}$ is planned by performing a "quantum free fall" experiment by detecting quantum interference of $\Hbar$ due to quantum gravitational bound states above a reflecting surface. Additionally, a $\Hbar$ Lamb shift measurement is being prepared in parallel that will allow to determine the antiproton charge radius at the level of $10\,\%$.}

\section{The GBAR experiment}
Neutral antihydrogen atoms ($\Hbar$), consisting of one antiproton ($\pbar$) and one positron ($\positron$), are used to test the behaviour of antimatter in the terrestrial gravitational field at the AD facility at CERN. A first direct approach by releasing $\Hbar$ from a magnetic trap constraints the ratio of the gravitational to the inertial mass of $\Hbar$ between $-65$ and $100$, with the sign of the gravitational acceleration $\gbar$ undetermined~\cite{Al_g}. The unique approach of the Gravitational Behaviour of Antihydrogen at Rest (GBAR) experiment is to synthesise positively charged antihydrogen ions ($\Hbarplus$) that can be trapped in a Paul trap and sympathetically cooled to a few $\mathrm{\mu K}$ temperature by coupling them to laser-cooled $\BeP$ ions, minimizing their initial velocities~\cite{GBAR_Prop}. The excess $\positron$ is then photo-detached using a $1640\,\mathrm{nm}$ laser and the now neutral $\Hbar$ experiences a classical free fall from a fixed height. By measuring the time of flight and annihilation position the acceleration $\gbar$ can be determined in a first phase with a precision of $1\%$. In a second phase measuring quantum interferences, due to quantum gravitational bound anti-atoms above a reflecting surface, will allow to reach a $10^{-6}$ precision~\cite{Crepin}. GBAR's unique formation of ultracold $\Hbar$ and $\Hbarplus$ may open the way for new exciting possibilities, such as optical trapping of ultracold anti-atoms~\cite{Crivelli_Trap} and the production of molecular anti-ions~\cite{Myers_MolIon}. Those experiments will result in an unprecedented sensitivity to test Lorentz and CPT symmetry, helping to shed light on the observed baryon asymmetry. 

As a byproduct of the free fall experiment, the $\Hbar$ Lamb shift, i.e. the energy difference of the $2\mathrm{S_{1/2}}$ and $2\mathrm{P_{1/2}}$ states, can be measured. Measuring the Lamb shift with an uncertainty of $100\,\mathrm{ppm}$ will determine the antiproton charge radius at the level of $10\%$ and probe CPT by looking for a difference in the energy levels from ordinary hydrogen atoms~\cite{Crivelli_PRD}.

\section{Production of antihydrogen atoms and ions}\label{sec:prod}
The $\Hbarplus$ is fabricated inside a so-called reaction chamber via two charge-exchange processes 
\begin{equation}\label{eq:Hbar}
    \begin{aligned}
    \pbar + \Ps &\rightarrow \Hbar + \electron \\
    \Hbar + \Ps &\rightarrow \Hbarplus + \electron.
    \end{aligned}
\end{equation}

The upgrade to the antiproton decelerator at CERN, the Extra Low ENergy Antiproton (ELENA) storage ring delivers every $110\,\mathrm{s}$ a bunch of about $5 \times 10^6$ $\pbar$ at a kinetic energy of $100\,\keV$. They have to be efficiently decelerated to a few $\keV$ energies for an optimal $\Hbar$ production. This is done using a drift tube that can switch from $-100\,\mathrm{kV}$ to $0\,\mathrm{V}$ when the $\pbar$ bunch is inside the tube. With this technique it was shown that the beam can be decelerated to $8\,\keV$, though it increases the beam emittance~\cite{Decel}.

\begin{figure}
    \centering
    \includegraphics[width=0.9\textwidth]{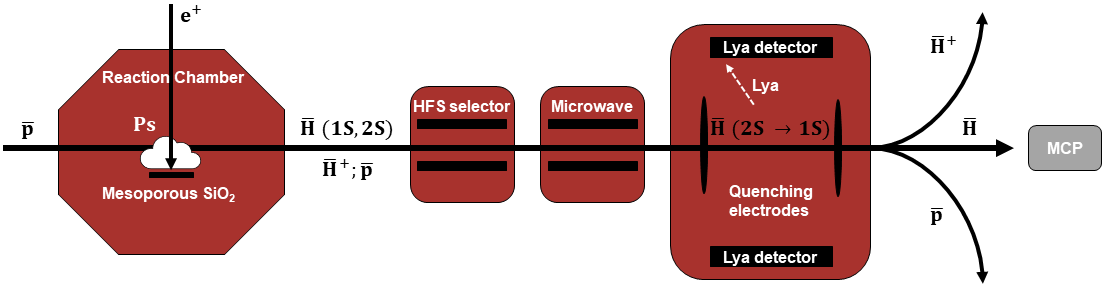}
    \caption{General scheme of the $\Hbar$ and $\Hbarplus$ production and the experimental setup for the $\Hbar$ Lamb shift measurement.}
    \label{fig:Setup}
\end{figure}

The GBAR LINAC produces electrons with a kinetic energy of $9\,\MeV$ in a pulse width duration of $2.85\,\mathrm{\mu s}$ and a repetition rate of up to $300\,\mathrm{Hz}$. Positrons are then generated via pair production of high-energy $\gamma$'s due to electrons impinging on a tungsten target. At a LINAC rate of $200\,\mathrm{Hz}$ the positron flux is about $4 \times 10^7\, \positron/\mathrm{s}$~\cite{PositronProd}. The $\positron$ are then accumulated in a buffer gas trap, and afterwards stacked in a ultra-high vacuum high field trap. Routinely trapping $1.5 \times 10^8\,\positron / 110\,\mathrm{s}$ has been achieved in the fall of 2021 and a maximum of $1.4 \times 10^9\,\positron$ within $1100\,\mathrm{s}$ has been demonstrated. The accumulated $\positron$ are then ejected onto a porous silica film inside a cavity, forming a positronium ($\Ps$) cloud in vacuum. The conversion probability of $\positron$ to ortho-$\Ps$ with a lifetime of $142\,\mathrm{ns}$ is about $30\,\%$. Due to losses in the $\positron$ transport onto the target about $1 \times 10^6\,\Ps$ were formed during an ELENA cycle inside the cavity of size $2\,\mathrm{mm} \times 2\,\mathrm{mm} \times 20\,\mathrm{mm}$  in 2021. The low energy $\pbar$ flying through the $\Ps$ cloud will first form $\Hbar$ and then $\Hbarplus$ as described in Eq.~\ref{eq:Hbar}. It follows that a high density of the order of $10^{11}\,\Ps/\,\mathrm{cm^3}$ is needed to form a single $\Hbarplus$~\cite{GBAR_Prop}. As a byproduct to the $\Hbarplus$ production many neutral $\Hbar$ atoms are formed and a parasitical Lamb shift measurement can be undertaken. Thus, the resulting beam of $\pbar$, $\Hbar$ and $\Hbarplus$ exits the reaction chamber, passes through the Lamb shift setup and is afterwards split in an electrostatic field to dedicated detection systems, as is shown in Fig.~\ref{fig:Setup}. 

\section{Lamb Shift measurement as a byproduct of antihydrogen production}\label{sec:Lamb}
A fraction of the produced $\Hbar$ will be in the metastable $2\mathrm{S_{1/2}}$ state. When passing through two microwave field regions they can be excited to the $2\mathrm{P_{1/2}}$ state depending on the resonance frequency. Since the transition frequencies of the hyperfine states are close to each other it is beneficial to remove unwanted states using a state selector and subsequently scan the isolated $2\mathrm{S_{1/2}}(F=0, m_F=0) \rightarrow 2\mathrm{P_{1/2}}(F=1, m_F=0)$ transitions with a MW transmission line, see Fig.~\ref{fig:Setup}. Remaining $\Hbar(2\mathrm{S})$ are quenched to the $2\mathrm{P}$ state in a static electric field where they relax to the ground state with a $1.6\,\mathrm{ns}$ lifetime emitting a Lyman-$\alpha$ photon with a wavelength of $122\,\mathrm{nm}$. With the surrounding CsI-coated MCP detectors the Ly-$\alpha$ photons are counted as a function of the MW frequency thus determining the $\Hbar$ Lamb shift. The setup was commissioned with Muonium at PSI where the detection efficiency was validated to be $\epsilon=\epsilon_{LyA}\times\epsilon_{\mathrm{Geo/Qnch}}=16\%$~\cite{MuBeam,MuLamb}. This includes the geometrical and quenching efficiency $\epsilon_{\mathrm{Geo/Qnch}}=40\%$ simulated using SIMION and the MCP efficiency with a quantum efficiency of $50\%$ to convert a photon to a free electron $\epsilon_{LyA}=\epsilon_{MCP}\times\epsilon_{QE}=40\%$~\cite{QE,MCP}.

\section{Capture and cooling of antihydrogen ions}\label{sec:Cooling}
After its production the $\Hbarplus$ is captured and cooled to ultracold temperatures in two steps. First, the single $\Hbarplus$ is trapped and sympathetically cooled by Coulomb interaction in a RF Paul trap filled with a laser-cooled $\BeP$ ion cloud. An efficient sympathetic cooling process is needed to circumvent a possible photo-detachment of the excess $\positron$ by the $313\,\mathrm{nm}$ laser used to cool the $\BeP$. It is most efficient when the mass ratio of the ions is small which is not the case for $\BeP$ and $\Hbarplus$ with a ratio of $9/1$. Simulations show that $\mathrm{HD^+}$ ions can be added to reach a temperature in the range of $5-100\,\mathrm{mK}$ within a cooling time of a few $\mathrm{ms}$~\cite{Cool}. To further cool the $\Hbarplus$ ion, it is then transported to a precision trap to form a $\BeP/\Hbarplus$ ion pair. Applying ground state Raman sideband cooling to the ion pair will cool the $\Hbarplus$ to $10\,\mathrm{\mu K}$ temperature corresponding to initial velocities of $\sim 1\,\mathrm{m/s}$~\cite{GBAR_Prop,Cool}.

\section{Classical free fall}\label{sec:Classic}
\begin{figure}[t]
    \centering
    \begin{minipage}{0.45\textwidth}
        \includegraphics[width=\textwidth]{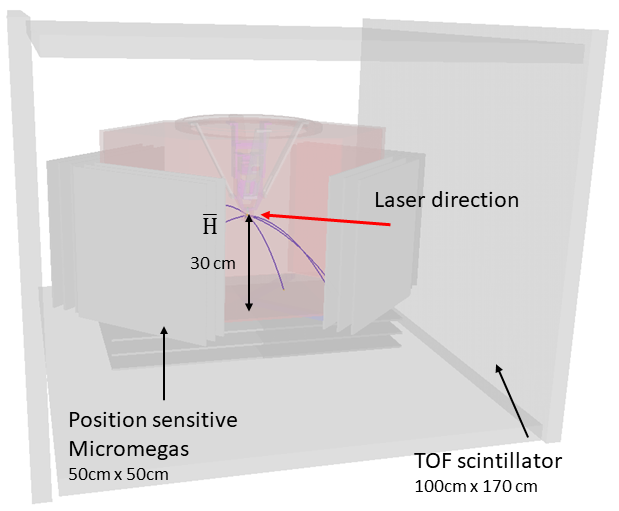}
        \caption{GEANT4 rendering of the classical free fall simulation with the tracks of the $\Hbar$ atoms (blue) inside the free fall chamber (red). The 6 Micromegas triplets and 4 TOF detectors are shown in grey.}
        \label{fig:FFC}
    \end{minipage}
    \hfill
    \begin{minipage}{0.45\textwidth}
        \includegraphics[width=\textwidth]{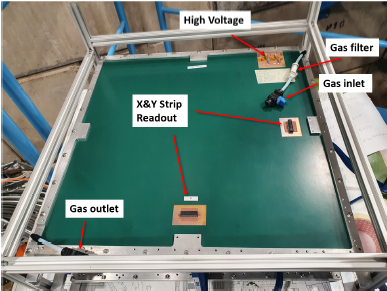}
        \caption{Micromegas detector used for tracking of charged pions from $\Hbar$ annihilations in the GBAR experiment~\protect\cite{Janka_Thesis}. \\}
        \label{fig:MM}
    \end{minipage}
\end{figure}
A laser pulse with a $1640\,\mathrm{nm}$ wavelength ionizes the $\Hbarplus$, triggering the start of the free fall $t_i$. The now neutral $\Hbar$ is no longer held by the precision trap and falls from a height of $z_i = 30\,\mathrm{cm}$ inside a so-called Free Fall Chamber (FFC), see Fig.~\ref{fig:FFC}. The initial velocity of the anti-atom directly affects its trajectory and recent simulations have shown that a hexagonal FFC maximizes the vertical flight path while having the best mechanical stability. The $\Hbar$ will inevitably come in contact with the vacuum vessel and annihilate, producing on average three charged pions $\pi^{\pm}$. Surrounding position sensitive Micromegas detectors track the trajectory of the pions, allowing for the reconstruction of the $\Hbar$ annihilation position with a focus on the vertical $z_f$-component. The Micromegas detectors of size $50\,\mathrm{cm} \times 50\,\mathrm{cm}$ are arranged in groups of three, of which at least two must record a pion hit to reconstruct its flight path~\cite{Janka_Thesis}. Additionally, TOF scintillator bars are forming the outer layer of the detection setup, and are used to measure the precise free fall time $t_f$ and help reject cosmic ray signals~\cite{TOF}. The gravitational acceleration is then extracted from the classical formula
\begin{equation}
    z_f = z_i + v_z^i(t_f-t_i) + \frac{1}{2}\gbar(t_f-t_i)^2
\end{equation}
where the initial vertical velocity $v_z^i$ is a priori unknown. Detailed simulations of the initial position and velocity distributions are ongoing and using a realistic GEANT4 simulation an advanced analysis of the reconstruction is studied.

\section{Quantum reflection and the quantum free fall}\label{sec:Quantum}
\begin{figure}[t]
    \centering
    \begin{minipage}{0.45\textwidth}
        \includegraphics[width=\textwidth]{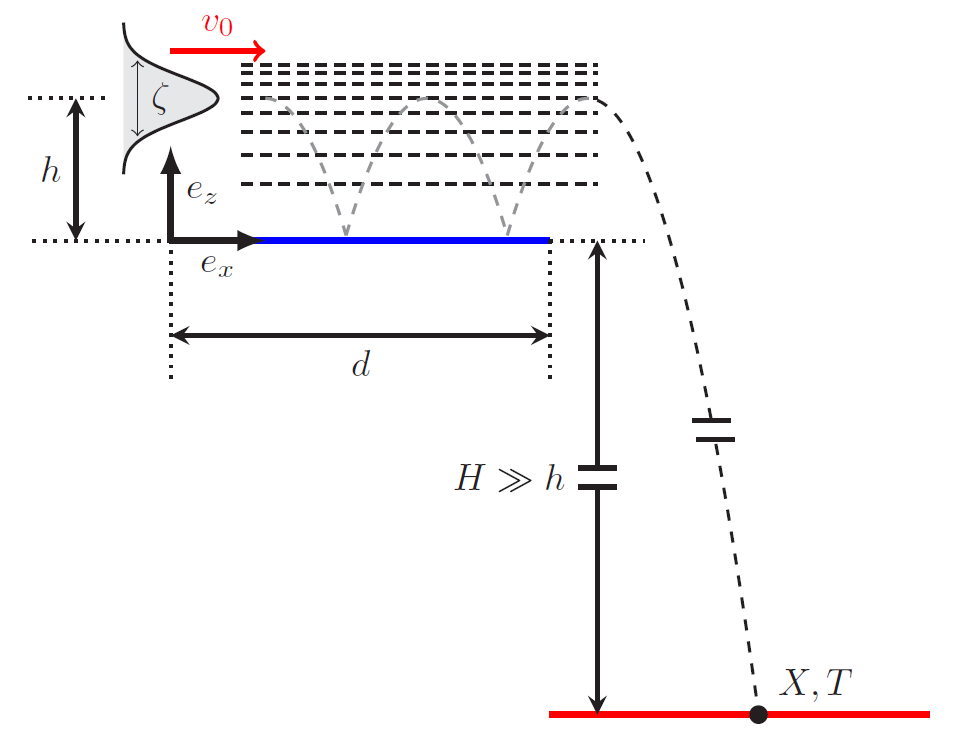}
        \caption{Two dimensional scheme for the detection (red line) of quantum interference due to quantum reflection of $\Hbar$ on a mirror surface (blue line)~\protect\cite{Crepin}.}
        \label{fig:2D}
    \end{minipage}
    \hfill
    \begin{minipage}{0.45\textwidth}
        \includegraphics[width=\textwidth]{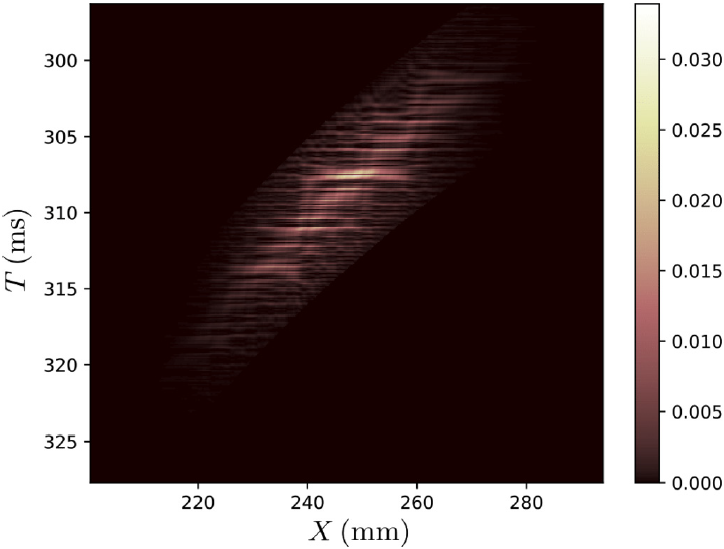}
        \caption{Probability distribution on the detection plane of the interference pattern in one dimension~\protect\cite{Crepin}. \\}
        \label{fig:Current}
    \end{minipage}
\end{figure}

The phenomenon of quantum reflection of $\Hbar$ on a surface rather than being annihilated, plays an important role for the GBAR experiment~\cite{Dufour}. It has to be taken into consideration in the analysis of the classical free fall experiment since $\Hbar$ can be reflected on the free fall chamber and thus increase the measured time of flight. But the effect can also be exploited to reach a $\gbar$ precision of $10^{-6}$ by detecting the interference pattern produced by anti-atoms bouncing on a mirror surface before experiencing a classical free fall~\cite{Crepin,Dufour}. The experimental idea is based on the concept of the neutron whispering gallery experiment~\cite{Neutron}. A sketch of this simplified scheme considering only two dimensions (X and Z) is shown in Fig.~\ref{fig:2D}. The $\Hbar$ atoms are dropped from a small height and bounce on a mirror of length $d$ and height $H$ above the detection plane. The dashed lines represent the $\Hbar$ quantum states due to the gravitational and mirror potential. The different possible paths through these states create an interference pattern in momentum space. By letting the atoms free fall from the end of the mirror results in a detectable interference pattern in position/time space as shown in Fig.~\ref{fig:Current}.

\section{Summary}\label{sec:Conc}
The unique approach of the GBAR experiment to measure the free fall of $\Hbar$ in the Earth's gravitational field, consists of synthesizing $\Hbarplus$ and cooling it to a few $\mathrm{\mu K}$ temperature, before photo-detaching the excess positron and releasing the neutral anti-atom. It has been shown that the individual $\pbar$ deceleration works as proposed. However, reliably producing, trapping and transporting positrons and consecutively forming $\Ps$ must be improved for the beamtime in $2022$ to get $\Hbar$ and $\Hbarplus$ in a second step. In parallel the $\Hbar$ Lamb shift setup has been commissioned with hydrogen and is ready for measurements with the anti-atoms as soon as those are available in GBAR. Detailed simulations of the classical and quantum free fall have been undertaken. 

\section*{Acknowledgments}
This work is supported by the Swiss National Foundation under the grants 197346 and 201465.

\end{document}